\documentclass[12pt]{iopart}
\usepackage{graphicx}
\usepackage{subfigure}
\usepackage{cite}
\usepackage{iopams}  
\usepackage{hyperref}
\begin{document}
\title[A novel liquid metal divertor with pre-filled microstructures]{Experimental investigation of a novel liquid metal plasma facing component with pre-filled microstructures}

\author{Yi-Jun Wang$^1$$^*$, Kai-Lun Li$^2$$^*$, Rui-Zhi Chen$^1$, Yue-Bin Hu$^1$, Juan-Cheng Yang$^3$, Ming-Jiu Ni$^{1,4}$\textsuperscript{\textdagger} and Zhao-Hui Yao$^{1,4}$\textsuperscript{\textdagger}}

\address{$^1$ School of Engineering Science, University of Chinese Academy of Sciences, Beijing 101408, China}
\address{$^2$ Institute of Engineering Thermophysics, Chinese Academy of Sciences, Beijing 100190, China}
\address{$^3$ State Key Laboratory for Strength and Vibration of Mechanical Structures, School of Aerospace,
Xi’an Jiaotong University, Xi’an, Shaanxi 710049, China}
\address{$^4$ State Key Laboratory of Nonlinear Mechanics, Institute of Mechanics and University of Chinese Academy of Sciences, Beijing 101408, China}
\address{$^*$These authors contributed equally.}

\ead{\mailto{mjni@ucas.ac.cn}, \mailto{yaozh@ucas.edu.cn} }

\begin{abstract}
Regarding the plasma facing components (PFCs) in nuclear fusion, liquid metal PFCs with stable free surface flow on PFC surface are considered a promising alternative. However, due to the poor wettability of liquid metal on most solid substrates and the complex magnetohydrodynamic (MHD), the realization of stable free surface flow on PFCs surface is challenging. In the present study, using the 3D printed methods, we developed a novel liquid metal PFC surface with MIcrostructures pre-FIlled by Liquid Metal (MIFILM) to realize a stable free liquid metal surface flow. The experimental results demonstrated that due to the existence of MIFILM, the apparent contact angle (ACA) of liquid metal changes from 140° to approximately 20°, indicating a transition from hydrophobic to hydrophilic. When the liquid metal flows on the MIFILM substrate, it is found that the liquid metal can completely spread on the surface with a stable and orderly free surface, even at a low flow rate. Moreover, the liquid metal could exhibit sustained spreading properties on the MIFILM substrate under a strong transverse magnetic field (up to 1.6 T). Results indicate that the magnetic field induces limited MHD drag but also accelerates the flow via two-dimensional effects. When the Stuart number N $<$ 1, the flow accelerates and the film thickness decreases. For N $>$ 1, both flow velocity and film thickness gradually stabilize. Therefore, the present novel MIFILM can offer a good choice for liquid metal PFC substrates in nuclear fusion.

\end{abstract}

\noindent{\it Keywords}: plasma-facing components, liquid metal, wettability, free surface
%

%
\maketitle


\section{Introduction}

The Tokamak device, a type of magnetic confinement nuclear fusion system, is internationally recognized as one of the most promising configurations for achieving nuclear fusion. The plasma-facing components (PFCs) serve as shields to protect other modules from extreme heat loads, neutron irradiation, and particle fluxes from the plasma \cite{federici2001plasma}. In recent decades, the concept of using liquid metals as PFC surface has been proposed \cite{ying2004exploratory,stubbers2005measurement}. By employing flowing liquid metal to cover critical areas, this approach aims to address the challenges posed by the short lifespan of solid materials. This attracts considerable attention to the liquid metal free surface flow in a strong magnetic field. However, a key premise for realizing the intended functions of a liquid metal PFC is that the liquid must spread uniformly over the substrate, covering all solid areas. Otherwise, the exposed solid substrate may still be at risk of damage. 

Achieving full coverage is challenging because of liquid metals' high surface tension and poor spreading characteristics. In addition, the strong magnetic field inside the tokamak further suppresses the spreading. Magnetic pressure can induce additional surface tension and increase the contact angle of liquid metal droplets \cite{sun2012strong}. For moving liquid, it cuts magnetic field lines and induces an opposing Lorentz force. vertical magnetic fields suppress the spreading area of droplets upon impact with solid surfaces \cite{zhang2016spreading}. Horizontal magnetic fields have a similar effect but also induce an anisotropic spreading pattern \cite{han2024maximum}. In the flow with a free surface of finite width, magnetohydrodynamics (MHD) resistance further leads to flow obstruction, an increase in film thickness \cite{yang2020magnetohydrodynamic}, the formation of surface jets \cite{meng2022experimental}, liquid metal detachment from the side walls \cite{narula2005study}, and even localized flow stagnation \cite{liu2018experimental}. This means that liquid metal film flow cannot fully cover the entire substrate under strong magnetic fields. So improving the wettability of liquid metal on conventional materials and generating stable thin film flow are urgent problems that needs to be addressed in the development of liquid metal PFCs.

The most straightforward approach to solve the problem is to increase the flow rate. The Reynolds number is typically used to characterize the relative magnitude of inertial forces in the flow, reflecting the impact of the flow rate. Studies on spreading characteristics indicate that even at a Reynolds number of 10\textsuperscript{4}, liquid gallium-indium-tin (GaInSn) metal flow exhibits a narrow rivulet-like pattern, where the flow detaches from the sidewalls, failing to spread fully. Stable coverage of the substrate material can only be achieved when the Reynolds number reaches above 20 000 (flow rate: 357.6 cm\textsuperscript{3}$/$s) \cite{yang2016flow}. At this point, the flow velocity is high, and the film's surface exhibits intense fluctuations in various directions, resulting in a chaotic turbulent state. Research also indicates that pure lithium can achieve complete spreading on a flat substrate at Reynolds numbers exceeding 3000. However, it fails to spread due to magnetic damping effects when exposed to a strong magnetic field \cite{liu2018experimental}. Additionally, the Experimental Advanced Superconducting Tokamak (EAST) deployed a direct current electromagnetic pump in its flowing liquid lithium limiter prototype experiment, which uses Lorentz forces to drive the lithium to a flow rate of 2 cm\textsuperscript{3}$/$s \cite{hu2016first}, only about 1\% of the flow rate required for stable spreading as mentioned in \cite{yang2016flow}. Therefore, spreading liquid metal by increasing the flow rate requires a high-capacity pump system, which poses engineering challenges.

The second approach involves utilizing capillary effects. At sufficiently high temperatures, liquid metals can wet common substrate materials, such as lithium on iron, molybdenum, and tungsten. Under the condition of wetting, it is proposed that capillary forces can act as a driving force for liquid flow. The Lithium–metal infused trenches (LiMIT) and the Flowing Liquid Lithium (FLiLi) divertor prototype adopt this principle, utilizing streamwise microchannels on the substrate to promote liquid metal coverage \cite{ruzic2011lithium, zuo2018investigation}. After four generations of improvements, it is estimated that the surface coverage rate of FLiLi has increased to 87\%, which is significant progress. However, flow traces shows that it still falls short of achieving ideal full coverage \cite{zuo2017mitigation,zuo2018results}. If certain streamwise channels do not contain liquid initially, it's difficult for the liquid in other channels to cross over the channel walls laterally and fill those empty channels, leaving them exposed. In the Magnum-PSI linear plasma generator, the solid substrate with stainless steel streamwise channel structures suffered localized melting when subjected to a plasma irradiation of 3 MW/m$^{2}$. This indicates that during the testing process, some areas of the substrate were exposed and not covered by lithium \cite{fiflis2015performance}.

The third approach involves surface treatments, such as laser ablation, high-temperature baking, glow discharge cleaning, and evaporated lithium coatings \cite{kietzig2009patterned, zuo2018investigation,fiflis2014wetting,krat2017wetting}. These methods improve the wettability of the surfaces, thus reducing the contact angle of the metal droplets on these surfaces and improving the spreading property. However, most of these treatments are one-time measures. After a certain period of use or system shutdown and restart, the surface needs to be retreated, as these methods do not permanently change the solid properties.

Moreover, with advancements in metal 3D-printing methods, capillary porous systems (CPS) with complex capillary channels have been proposed and tested in various plasma devices \cite{yuan2024design,rindt2021conceptual,scholte2023design,rindt2021performance,rindt2019using}. In comparing experiments with tungsten meshes, tungsten fiber mats, and tungsten sintered disks, tungsten capillary plates have shown their effectiveness in withstanding plasma pulses and enduring higher heat flux densities without surface damage or particle ejection \cite{oyarzabal2023comparative}. However, it is important to note that CPS differs from the free surface film flow system. The CPS continuously pumps liquid outward through numerous holes dispersed across the backing plate. Driven by capillary forces, liquid metal climbs within the channels and spreads outwards upon reaching the upper surface of the plate to wet its surrounding area. If the capillary channels become obstructed by impurities from the filtering process, the CPS will lose its protective layer of liquid metal.

As a novel method for spreading free surface film flow, this study prepares a substrate with a triangular array of thumbtack-shaped micropillars using the 3D-printing method and pre-filled with liquid metal. The solid-liquid combined surface enhances the wettability of the liquid metal on the substrate material, reducing the apparent contact angle (ACA) and enabling the liquid to wet the substrate material and spread automatically. This approach significantly reduces the minimum temperature (down to the melting point of the liquid metal) and the flow rate (down to \textit{Re} $\sim$ 10\textsuperscript{2}) required for spreading. Compared to CPS, the pre-filled substrate serves primarily as a flow substrate that improves wettability. During operation, the mainstream of liquid metal flows over the substrate, from one end to the other, rather than exiting through the complex internal channels. This design eliminates concerns about channel clogging due to impurities or solid damage. 

Section \ref{s2} presents the design and fabrication of substrates with MIcrostructures pre-FIlled by Liquid Metal (MIFILM), and their experimental effect in reducing the contact angle and promoting wettability. Based on this, liquid metal free-surface flow on MIFILM substrates will be discussed in section \ref{s3}. A transverse magnetic field is applied. It aims to mimic the magnetic environment of the divertor in a Tokamak, proving the preliminary feasibility of the proposed design as a novel PFC. Finally, section \ref{s4} summarizes the results of this study and outlines future directions for further research.

\section{Wetting liquid metals on MIFILM substrates} \label{s2}

\subsection{Design and fabrication of microstructured substrate}

As early as 1805, Young noted that the contact angle on an ideally smooth surface is determined by the physical properties at the three-phase contact line \cite{young1805iii}. It can be obtained according to the balance of horizontal forces when thermodynamic equilibrium is reached. Subsequently, Wenzel \cite{wenzel1936resistance,wenzel1949surface} and Cassie \cite{cassie1944wettability,cassie1948contact} explained the role of microstructures in enhancing intrinsic wettability. In the Wenzel state, a wetting droplet exhibits increased wettability and a smaller contact angle due to surface roughness. In contrast, in the Cassie state, a non-wetting droplet has reduced wettability and a larger contact angle due to the surface structures.

Driven by urgent demands such as waterproofing, anti-icing, and self-cleaning, plenty of research has been conducted on the transition process from wetting to non-wetting \cite{bush2006walking,gao2004water}. Without relying on chemical modifications, inspired by biological waterproof surfaces, reentry microcolumns with overhangs can make the surface repel liquids with contact angles less than 90 degrees. Meanwhile, doubly reentry microcolumns with secondary overhanging structures can repel liquids even when their contact angles are close to 0 \cite{liu2014turning,hensel2013wetting}.

Conversely, due to a lack of application scenarios, research on the transition of surfaces from non-wetting to wetting has received less attention. However, a recent experimental study showed that mercury can wet along pre-filled reentry microchannels. Its ACA is less than 30 degrees in the direction of the channel with an intrinsic contact angle of more than 140 degrees \cite{wilke2022turning}. Simple square microchannels with no reentry can only enhance the non-wettability of droplets, and the liquid does not stay in the gaps. The main difference lies in the design of the reentry structures, which aim to create an energy barrier through their overhanging structures. The surface energy is minimized when the liquid precisely fills the gaps between the microstructures. Any further increase or decrease in the liquid volume will raise the system’s energy, thus maintaining the filled state as a quasi-steady state and ensuring that non-wetting liquids do not escape.

To spread the liquid metal on substrates such as divertor surfaces, it is essential to further advance the work in \cite{wilke2022turning}, in which only one-dimensional linear wettability in trenches was reached. In this paper, we will extend the linear wettability to full-direction wettability on the two-dimensional surface, and droplet wettability to film flow spreadability, by designing a triangular lattice arrangement of thumbtack-shaped micropillars.

\begin{figure}
\centering
\includegraphics[width=0.8\linewidth]{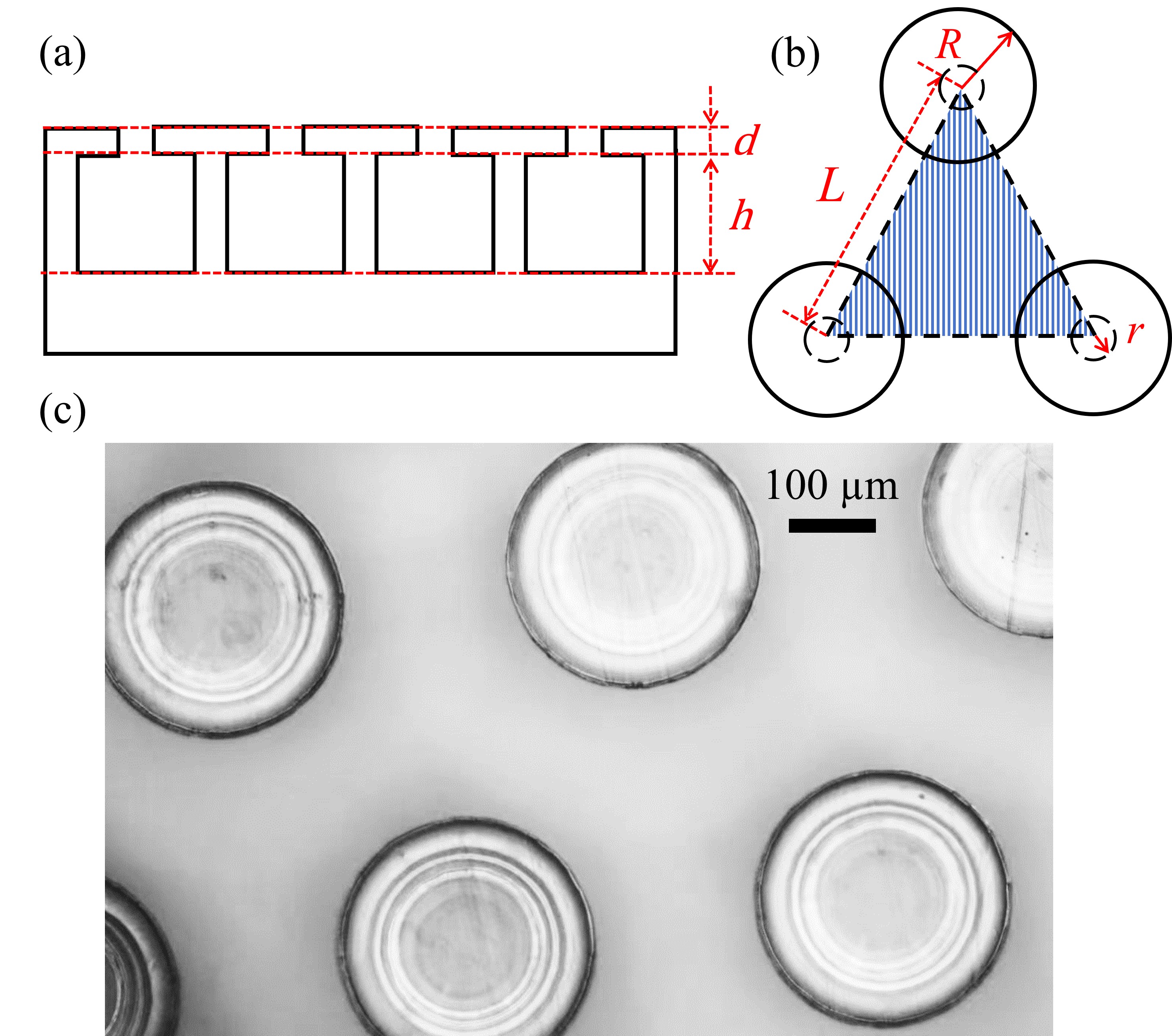}
\caption{Design of the microstructures: (a) The side view schematic and (b) the top view schematic of the thumbtack-shaped micropillar array. (c) A top view of the micropillars' round caps. }
\label{f1}
\end{figure}

This study utilizes photocurable 3D-printing technology (BMF NanoArch S130, Phrozen Mighty 8K) to fabricate microstructured substrates. Photosensitive resin provides higher resolution as a substrate material. The planar resolution of the photocuring process varies from 2 to 20 µm, with a fixed layer thickness of 5 µm. The base version of this triangular-arranged micropillar array is initially designed with the radius \textit{r} = 60 µm and the height \textit{h} = 600 µm of the pillar, the radius \textit{R} = 150 µm and the thickness \textit{d} = 15 µm of the top overhang, and the spacing between two pillars nearby \textit{L} = 500 µm, as shown in \fref{f1}(a) and (b). The solid fraction \textit{f} refers to the proportion of the top solid area to the projected area of the surface, represented here as the ratio of the shaded sectors and the whole shaded triangular area in \fref{f1}(b). Below the top surface, all sharp edges are rounded to facilitate gas removal. Upon 200$\times$ magnified observation, the micropillars show no significant serration defects, as illustrated in \fref{f1}(c). Two strategies are utilized to alter the relevant sizes of the microstructure and quantitatively study their effects: one involves keeping other parameters constant while varying the spacing \textit{d}, which affects the solid area fraction \textit{f}. Another approach is to fix the relative proportions of all dimensional sizes, scaling the element (the shaded area in \fref{f1}(b)) as a whole to observe the effects of the change in characteristic length. A total of 48 square structured substrates are printed, with their \textit{f }ranging from approximately 0.08 to 0.34, and \textit{R} of 150, 300, 350, and 450 µm. 

In our research, the largest pillar design which still achieves the desired wetting function, is finally selected for its engineering practicality, as it shows higher structural strength and large-area manufacturing feasibility. These larger structures will serve as the benchmark for the design of subsequent long film flow substrates in section \ref{s3}.

\subsection{Substrate pre-filling}

\begin{figure}
\centering
\includegraphics[width=0.8\linewidth]{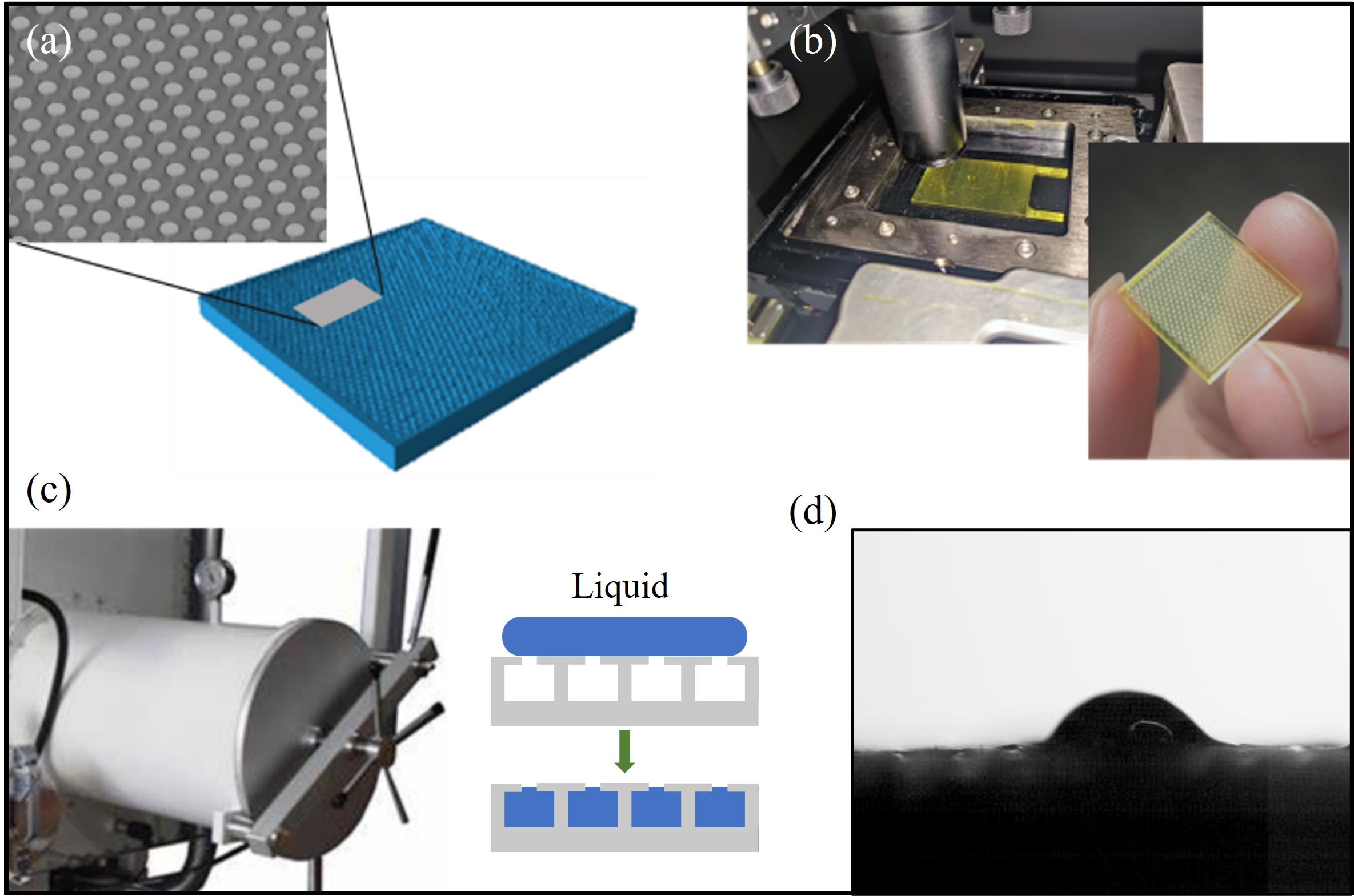}
\caption{Wettability adjustment experiment process: (a) CAD modeling, (b) photopolymerization 3D-printing, (c) vacuum injection of liquid metal, (d) measurement of contact angle for test samples. }
\label{f2}
\end{figure}

After computer-aided design (CAD) modeling and 3D-printing of the microstructured surface (shown in \fref{f2}(a) and (b)), the liquid metal is introduced into the structure gaps using a vacuum injection method. The liquid metal employed in this experiment is a GaInSn alloy. The alloy is silvery in color and remains liquid at room temperature. GaInSn is non-toxic and harmless, and its surface tension coefficient, intrinsic contact angle on solid substrates, and electrical conductivity are close to those of liquid lithium, making it a good substitute for lithium in flow experiments. Specifically, the alloy used in the experiment is Ga\textsuperscript{67}In\textsuperscript{20.5}Sn\textsuperscript{12.5}, and its density $\rho$ = 6360 kg/m\textsuperscript{3}, surface tension $\gamma =0.533$ N/m, kinematic viscosity $\nu =2.98\times {{10}^{-7}}$ m\textsuperscript{2}/s, and electrical conductivity $\sigma$~ = 3.1$\times {{10}^{6}}$ $\Omega$\textsuperscript{-1}m\textsuperscript{-1} \cite{morley2008gainsn}. The filling process of the structured substrate is conducted entirely under an argon atmosphere within a glovebox, with an oxygen content of less than 2 ppm.

The vacuum injection method has been widely adopted in the assembly of lab-on-chip devices, fabrication of liquid metal micro-electromechanical systems (MEMS), and printing of flexible circuits due to its ability to eliminate air bubble cavities in the channels, compared to the direct injection method \cite{fassler20133d,lin2017vacuum,andrews2018patterned}. Vacuum suction can provide a pressure difference of about 1 atmosphere. According to the prediction in \cite{zheng2005effects}, the critical pressure required for the suspended droplets in the Cassie state to “de-pin” and enter the gaps between the underlying microcolumns can be expressed as:
\begin{eqnarray}
    P_c=\frac{\gamma \cdot f\cdot \cos \theta }{(1-f)(A/L)} ,
\end{eqnarray}
where $\theta$ is the intrinsic contact angle, \textit{A} is the area of the microcolumn tops, and \textit{L} is the perimeter of the microcolumns, and the ratio of the two represents the scale of the microcolumn. Estimates indicate that a pressure \textit{P} $< $ 10$^4$ Pa is required to fill the substrate, and the vacuum injection method meets this pressure requirement. The semi-auto operational process for the pre-filling of the substrate is as follows: The substrate is fixed in an empty container and placed in a gas-tight chamber. A vacuum pump is operated continuously for approximately 30 minutes to remove gases, and the chamber pressure gauge shows -0.1 MPa. Then, a pair of electric push rods extends by remote control to pour liquid metal into the container. The liquid metal covers the substrate to be filled, and argon gas is pumped into the chamber to increase the pressure, pushing the liquid metal into the gaps among the micropillars shown in \fref{f2}(c). Then, another push rods lift the MIFILM substrate out of the liquid pool. The process can also be efficiently performed using robotic arms for future high-temperature lithium filling operations.

After the GaInSn is pre-filled, droplets of the same liquid are dripped onto the MIFILM substrate through a PEEK tube to measure the droplet morphology and ACA. The contact angle experiment is illustrated in \fref{f2}(d). 

\subsection{Enhanced wettability of liquid metal}

Young's equation derives the intrinsic contact angle $\theta$ of a liquid on an ideal surface from the balance of surface tension:
\begin{eqnarray}
     {{\sigma }_{gl}}\cos \theta ={{\sigma }_{sl}}-{{\sigma }_{sg}},
\end{eqnarray}
where ${{\sigma }_{gl}}$, ${{\sigma }_{sl}}$, and ${{\sigma }_{sg}}$ are the interfacial tension coefficients between the gas-liquid, solid-liquid, and solid-gas phases, respectively. The actual ACA of a composite surface composed of various material types can be described by the Cassie-Baxter equation:
\begin{eqnarray}
     \cos \theta ^*=\sum\limits_{i}{{{f}_{i}}}\cos {{\theta }_{i}} , 
\end{eqnarray}
where the overall ACA is given by $\theta ^*$, the area fraction of the \textit{i}-th surface material component is ${{f}_{i}}$, and the liquid's intrinsic contact angle on it is ${{\theta }_{i}}$.

\begin{figure}
\centering
\includegraphics[width=0.7\linewidth]{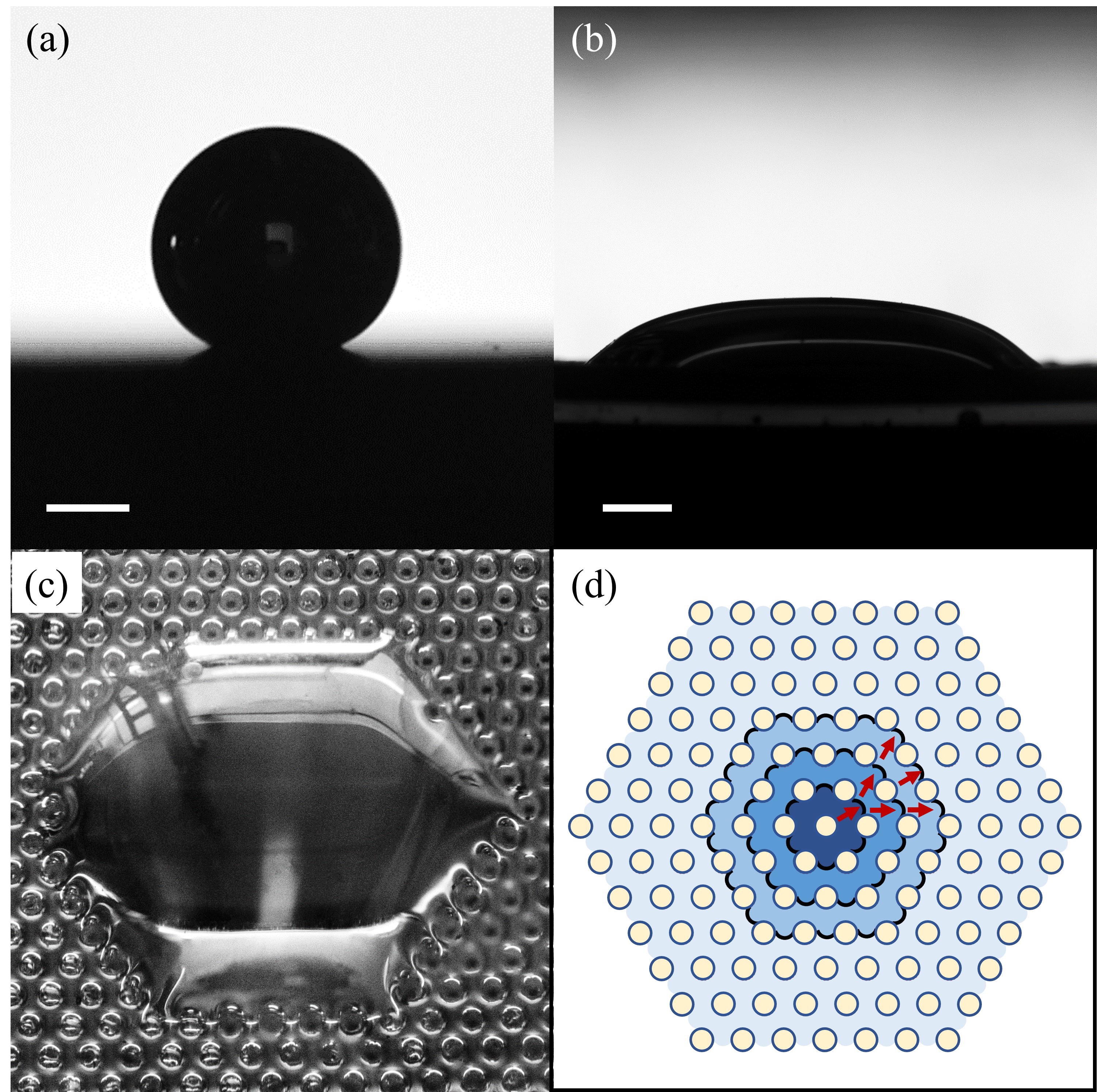}
\caption{Wetting behavior of GaInSn droplet on flat and MIFILM substrates. (a) A GaInSn droplet on the flat substrate shows a non-wetting state (contact angle 141.5°). (b) The droplet on the MIFILM substrate exhibits a wetting state (contact angle 40.6°), with a scale bar of 1 mm in both figures. (c) An oblique top view of the droplet on the MIFILM substrate. (d) A schematic illustration of the droplet spreading process (top view), the yellow circles represent micropillars. Assuming a droplet falls on the surface, the contact line advances outward from the initial contact point, gradually evolving into a hexagonal pattern. The black solid lines represent the contact line evolving with time, and the red arrows are the local advancing direction of the liquid.}
\label{f3}
\end{figure}

The main difference between the processes of “transitioning from non-wetting to wetting” and “transitioning from wetting to non-wetting” lies in whether the gaps between the microstructures are initially filled with the same liquid or gas. 
For a dry microstructure surface, the components of the substrate consist only of solids and air. Therefore, the equation simplifies to:
\begin{eqnarray}
   \cos \theta ^*=f_1\cos {\theta _1}+f_2\cdot (-1),{{f}_{1}}+{{f}_{2}}=1.
\end{eqnarray}

The effect of the gas portion of the surface is to increase the ACA. 
For MIFILM, the components of the substrate consist of the solid and the identical liquid itself:

\begin{equation}
    \cos \theta ^*={{f}_{1}}\cos {{\theta }_{1}}+{{f}_{2}}\cdot 1,{{f}_{1}}+{{f}_{2}}=1.
    \label{e5}
\end{equation}

The effect of the liquid portion is to decrease the ACA. Therefore, adjusting the liquid/solid area fraction of the composite surface can regulate the size of the final ACA when the other geometric parameters of the structured surface remain unchanged.

\begin{figure}
\centering
\includegraphics[width=0.7\linewidth]{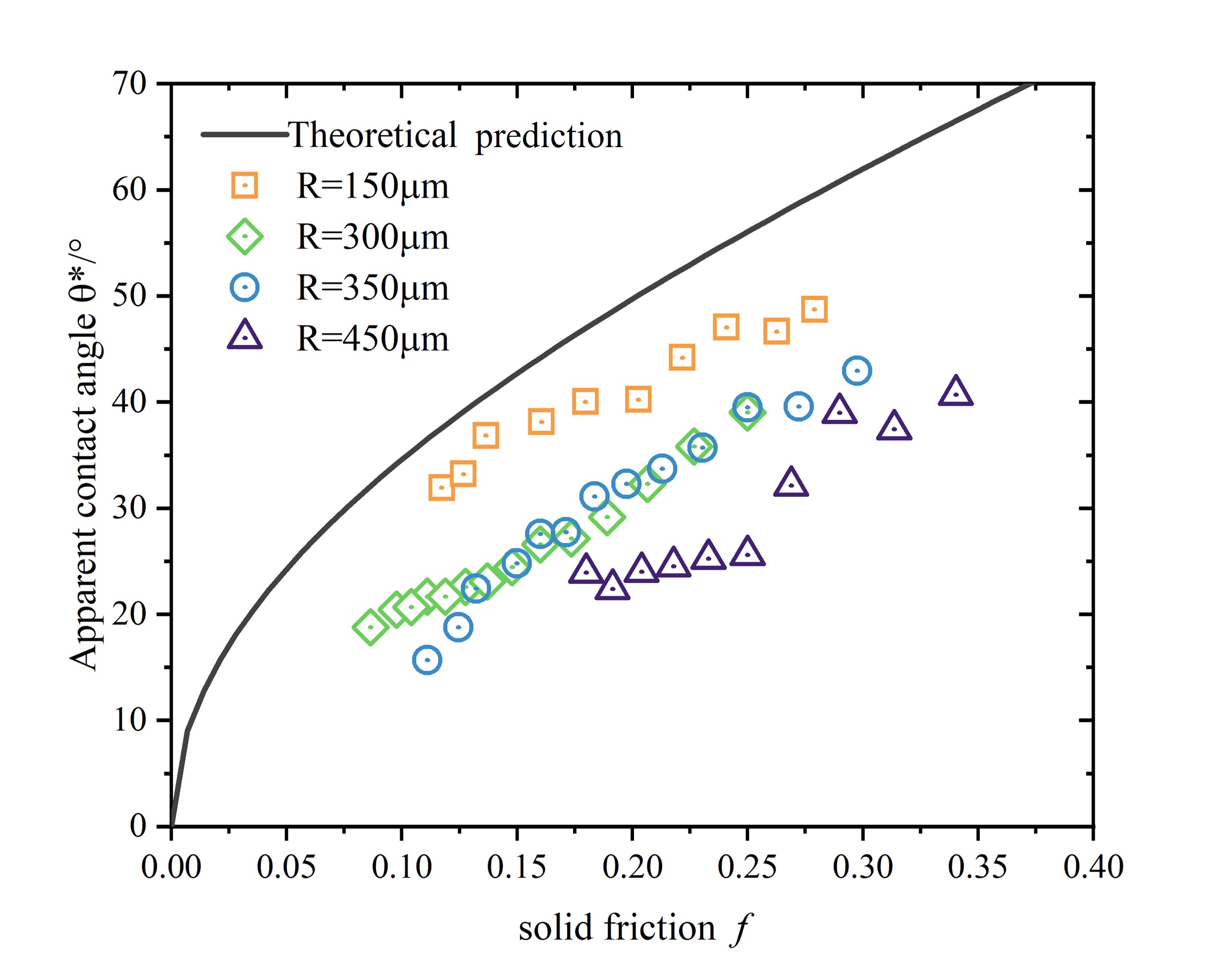}

\caption{The relationship between the solid friction \textit{f }and the apparent contact angle ${\theta ^*}$, where theoretical predictions are based on equation (\ref{e5}). \label{f4}}

\end{figure}

In \fref{f3}(a), a GaInSn droplet on the flat substrate shows a non-wetting state (contact angle 141.5°). \Fref{f3}(b) and (c) display side and top-view photographs of a droplet on the MIFILM substrates. It exhibits a wetting state (contact angle 40.6°). The unique hexagonal appearance of the droplet in \fref{f3}(c) arises from the anisotropy in the solid-liquid composition of the substrate. This phenomenon occurs when droplets simultaneously contact hydrophilic patterned surfaces \cite{courbin2007imbibition} and are forced to contact hydrophobic patterned surfaces \cite{lou2022polygonal}, which is primarily influenced by the arrangement and shape of the microstructures. 

\Fref{f3}(d) explains the formation mechanism of \fref{f3}(c) by illustrating how the contact line at the periphery of the droplet is hindered and pinned at the solid sections while continuing to extend forward at the liquid sections. As a droplet gradually falls onto the composite surface, its contact line advances outward in a “zipping” manner, moving from one layer of pillars to the next outer layer \cite{raj2014high}. In the micropillar array design presented in this study, the equilateral triangular arrangement results in the formation of hexagonal droplets. 

The widest part of the droplet profile in the backlit photograph is a protruding leading edge of the liquid-liquid contact, which means that fitting the droplet profile to measure the contact angle inevitably overestimates the influence of the liquid. \Fref{f4} shows the contact angles of liquid metal on pre-filled composite substrates with varying pillar characteristic scales and solid area fractions. The observed trend in the ACA is consistent with the Cassie-Baxter model’s predictions for solid-liquid composite wetting behavior. However, the actual liquid-solid composite surfaces are not ideally isotropic, resulting in measured ACA smaller than expected. Furthermore, as the characteristic scale of the microstructure increases, the effects of the protruding leading edge from liquid-liquid contact and the inward pinning at the contact line due to solid-liquid interactions become more obvious, leading to greater deviations.

\section{Free surface flow on MIFILM substrates}\label{s3}

\subsection{Setup of the flow experiment}

\begin{figure*}
\includegraphics[width=1\linewidth]{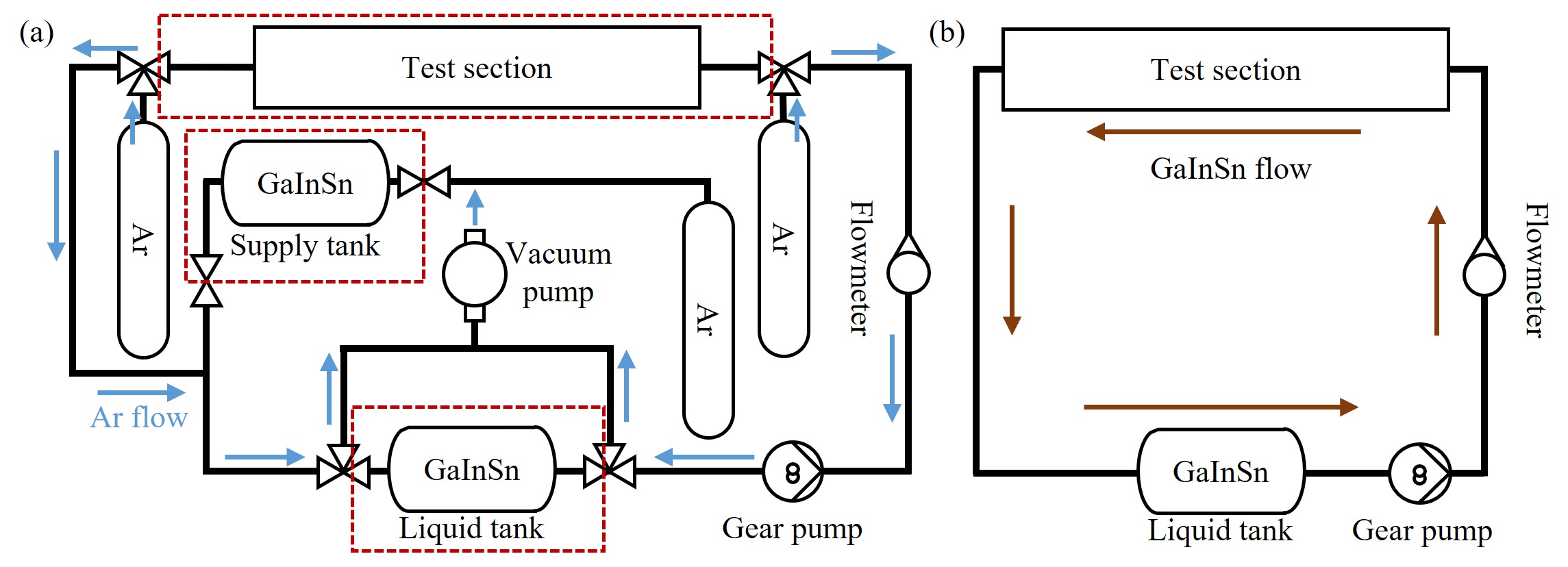}
\caption{Diagram of the experimental loop system: (a) Ar loop as the anti-oxidation measure. (b) The sealed liquid metal loop. }\label{f5}
\end{figure*}

The experimental system used for free surface flow experiments includes the liquid metal loop, Ar gas loop, electromagnet, and measurement devices. \Fref{f5} illustrates schematic diagrams of the liquid metal and gas loops consisting of the test section, a gear pump (Leapfluid CT3000F), a flow meter, a circulating liquid tank, a supply liquid tank, valves, three-way valves, a vacuum pump, argon gas sources, and pipelines.

The oxidation of liquid metal can significantly affect its physical properties, increasing viscosity and reducing surface tension. When the liquid metal recedes or its volume decreases, the oxide layer remains adhered in place. It is reported that the metal droplet gradually collapses until the measured receding angle is less than 20 degrees, at which point the oxide skin finally ruptures \cite{joshipura2021contact}. The “fake spreading” achieved through the adhesion of the oxide layer does not demonstrate that pure liquid metal can cover the entire substrate under the same conditions. To minimize this interference, the following anti-oxidation measures are taken. 

The main components within the red dashed box in \fref{f5}(a) are assembled under argon protection. Freshly prepared GaInSn liquid is filled in the tanks, and a MIFILM substrate is positioned inside the test section. The assemblies are then sealed and transferred to the external atmosphere. After the pipelines are connected but before the valves are opened, the pipelines are treated by connecting one end of each pipeline to a vacuum pump and the other to an argon source, which is shown in \fref{f5}(a) by the blue arrows. Drawing on the operating procedures of a glovebox, the pipelines are evacuated and pressurized three times; after that, they can be considered an oxygen-free environment. Finally, the three-way valves are switched to connect the pipelines, the test section, and the GaInSn tank to make the experimental loop a unified whole. The argon gas pressure is also used to drive GaInSn from a supply tank to the loop. When the pipeline is filled with liquid, the valve to the supply tank is closed. At this point, the internal environment is sealed, maintained at an argon-positive pressure, and ready for the flow experiment driven by a pump, as shown in \fref{f5}(b). Each experiment runs for no more than one hour to avoid oxidation. After that, the liquid is drained, the components are cleaned, and then reassembled for the next experiment. 

Once pumped from the liquid tank into the test section, the liquid metal first passes through the film flow generator and then flows over the test substrate. The film flow generator is placed at the entrance of the test section and consists of a series of 2 mm high, fully separated channel openings that span the entire width of the substrate, converting the pipeline flow into film flow. The area for the free surface film flow is 200 mm × 60 mm in size, which is mainly limited by the internal space of the electromagnet. The flow substrate is interchangeable, enabling tests with both MIFILM and flat surfaces. At the downstream end of the test section, the free surface flow falls along a stepped cliff to reproduce pressure conditions at infinite distances and prevent pipe backflow. 

\begin{figure*}
\includegraphics[width=1\linewidth]{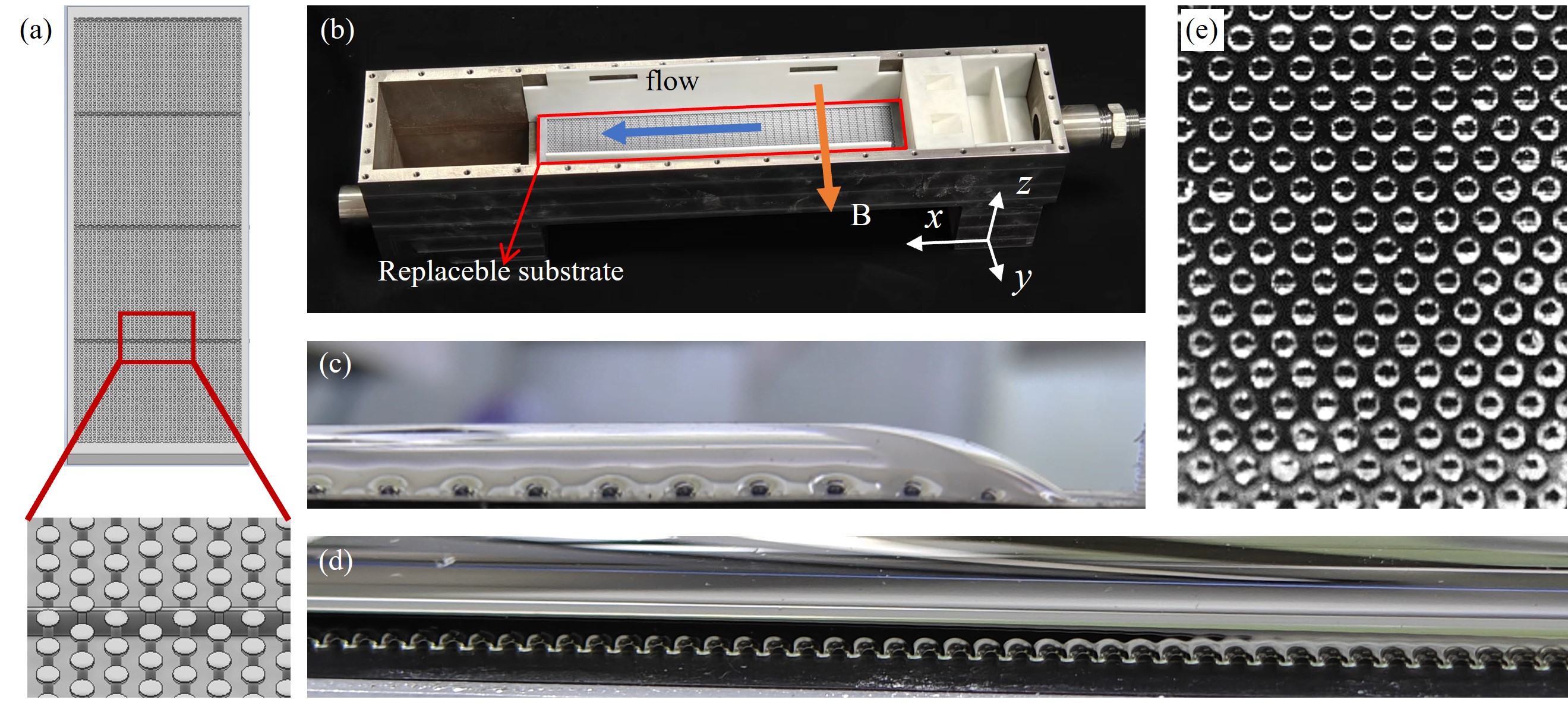}
\caption{Design, installation, and detailed photos of the long substrate for free surface flow experiments. (a) Under-liquid microwalls for better stabilization. (b) A photo of the test section, where the central area can accommodate either a flat or a MIFILM substrate for testing. (c) Side view of the liquid film spreading on the MIFILM substrate, advancing at an angle of approximately 37.3°. (d) Top view of the filling edge of the MIFILM. The clear and sharp contact line indicates that the filling is not a result of oxide skin adhesion. (e) Top view of the MIFILM surface.\label{f6}
}
\end{figure*}

Based on the results in section \ref{s2}, this study selects the largest micropillar size that reduces the ACA, specifically thumbtack-shaped micropillars with a top radius of 450 µm and a height of 800 µm. These micropillars result in an ACA of approximately 37° for GaInSn liquid. In addition, due to the large area of the MIFILM for the flow experiment, we installed microwalls (shown in \fref{f6}(a)) every 50 mm to segment the pre-filled liquid metal into four liquid pools. This design helps to block disturbances from propagating between pools, thereby enhancing the stability and tilt resistance of the pre-filled liquids. The assembled test-section module is shown in \fref{f6}(b), the acrylic cover removed. \Fref{f6}(c), (d), and (e) provide detailed views from different angles to display the static spreading of liquid metal on the MIFILM substrate. 

During the experiments, the test section is placed inside an electromagnet. The magnet can provide a transverse magnetic field ranging from 0 to 1.61 T by adjusting the input current. The test section is placed on a fixed bracket to ensure the free surface flow area is positioned within the uniform magnetic field region, which measures 80 mm × 200 mm.

The main dimensionless parameters in the experiment are the Reynolds number, the Hartmann number, and the Stuart number. The Reynolds number is defined as $Re={Q}/{\nu w}\;$, where \textit{Q} is the flow rate, and $w$ is the width of the flow. The Hartmann number is defined as $Ha=Bl \sqrt{{\sigma }/{\rho \nu }\;}$, where \textit{B} is the magnetic field strength, and the characteristic length \textit{l} is chosen as the initial film thickness at the inlet. The primary objective of this experiment is to examine whether the liquid film can achieve a stable spread on MIFILM substrate at significantly lower flow rates and to compare this flow with that on the flat substrate. Therefore, a lower initial Reynolds number is chosen for the experiment. The stability of metal liquid spreading on MIFILM substrate under different Reynolds numbers and Hartmann numbers is also studied. The experimental Reynolds number ranges from approximately 600 to 1500, and the Hartmann number varies from 0 to 110. Additionally, the magnetic interaction number or Stuart number $N={H{{a}^{2}}}/{Re}\;$ is used to compare the magnitudes of electromagnetic forces and inertial forces.

\subsection{Flow measurement methods
}
\begin{figure*}
\includegraphics[width=1\linewidth]{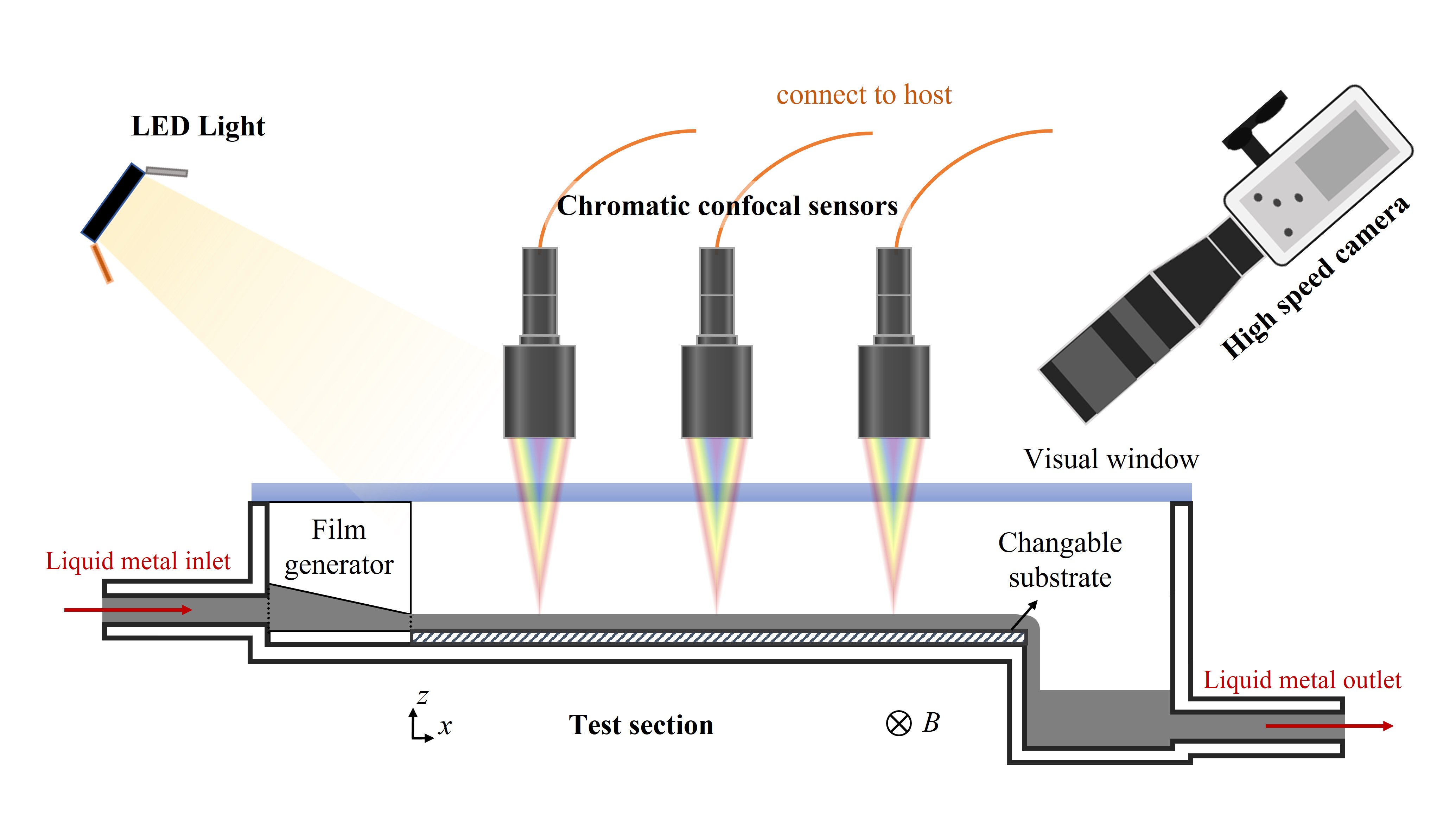}
\caption{A diagram illustrating the test section and the measurement methods employed in the present experiments.}\label{f7}
\end{figure*}

The test section features a stainless-steel body with a full-length acrylic top window for optical access. A high-speed camera (Photron NOVA S12) with a macro lens (Nikkor AF-S VR Micro 105mm f/2.8G IF ED) is used to observe flow patterns, and three chromatic confocal sensors (CCSs, Sincevision SC20011) provide more precise liquid film surface height variation information, as shown in \fref{f7}.

In previous experimental studies on the non-contact measurement of the flow characteristics of liquid metals, researchers used ultrasonic probes \cite{trifonov2017nonlinear,li2004experimental} or lasers \cite{yang2018rearrangement,yoshihashi2016laser} to measure the height of the flowing surface. However, in this experiment, the ultrasonic method is not adopted due to the influence of the complex solid-liquid interface on the substrate. The distance measurement of the laser profiler is based on triangulation, and the height of the reflection point is calculated according to the relative position of the reflection point and the emitter \cite{hausler1988light}. The receiver can easily detect light reflected in various directions from a diffuse surface. However, pure liquid metal has a smooth, mirror-like surface, and the receiver must be positioned at a specific angle (the optimal angle) relative to the light source to capture the reflected light. Therefore, previous experimental studies allowed for a small amount of oxidation, resulting in a thin oxide layer that converts specular reflection into diffuse reflection. This approach is incompatible with strict antioxidant measures. To address this issue, we adopted CCS. The dispersed white light can be seen as a series of laser beams of different colors, covering a broad range of angles. When a specific laser beam is reflected to the receiver, the receiver determines its path based on its wavelength, so CCS can measure a specular surface with unknown height and angle in a wider range. Compared with the laser profiler, CCS has better resistance to fluctuations.

Three CCSs are positioned upstream, midstream, and downstream of the flow (at distances of \textit{x} = 40 mm, 100 mm, and 160 mm from the film flow inlet, respectively). For each \textit{Re} and \textit{Ha} case, data is continuously collected at a frequency of 10 kHz for 20 seconds once the flow is stable.

\subsection{Liquid spreading performance}

\begin{figure}
\centering
\includegraphics[width=0.8\linewidth]{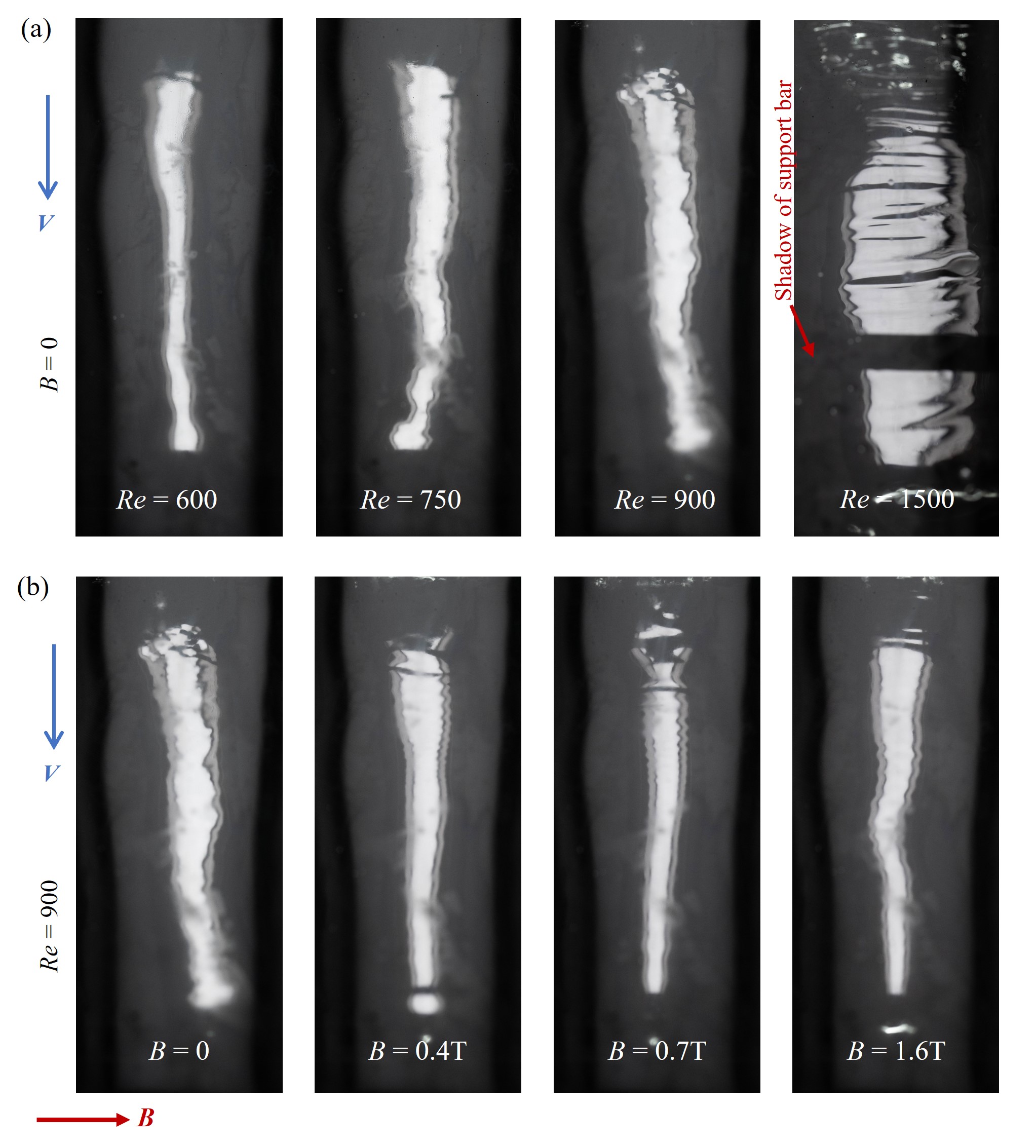}
\caption{Flow surface morphologies on the flat surface under different conditions. (a) The various \textit{Re} conditions with no magnetic field on the flat substrate, and (b) the different \textit{B} at \textit{Re} = 900 on the flat substrate.}\label{f8}
\end{figure}

\Fref{f8}(a) and (b) illustrate the flow behavior of the liquid metal on the flat surface under different Reynolds numbers and Hartmann numbers. On the flat surface, within the experimental flow range (\textit{Re} up to 1500), the liquid metal fails to fully cover the entire substrate area after passing through the film flow generator, but rapidly shrinks into a rivulet. The darkest black regions represent shadows from the sidewalls, the gray areas indicate exposed substrate, and the lighter areas are the GaInSn liquid. After implementing strict anti-oxidation measures, the GaInSn exhibits a high contact angle and a low sliding angle on the flat surface, resulting in randomness in the width and y-position of the rivulet, which can fluctuate across the span due to disturbances such as local roughness on the substrate. As the flow rate increases, the stream becomes wider, but it still does not make firm contact with the sidewalls. When a magnetic field is applied, the liquid metal flow becomes more stable, eliminating random displacements, while fluctuations in flow direction can be seen upstream of the free surface. We repeated the experiment in \cite{yang2016flow} and found that the liquid could fully spread only when the measured Reynolds number reached 18 500, which is consistent with the results of Yang \etal. The experiments indicate that at low flow rates, it is indeed challenging for liquid metal to spread over the substrate.

\begin{figure}
\centering
\includegraphics[width=0.8\linewidth]{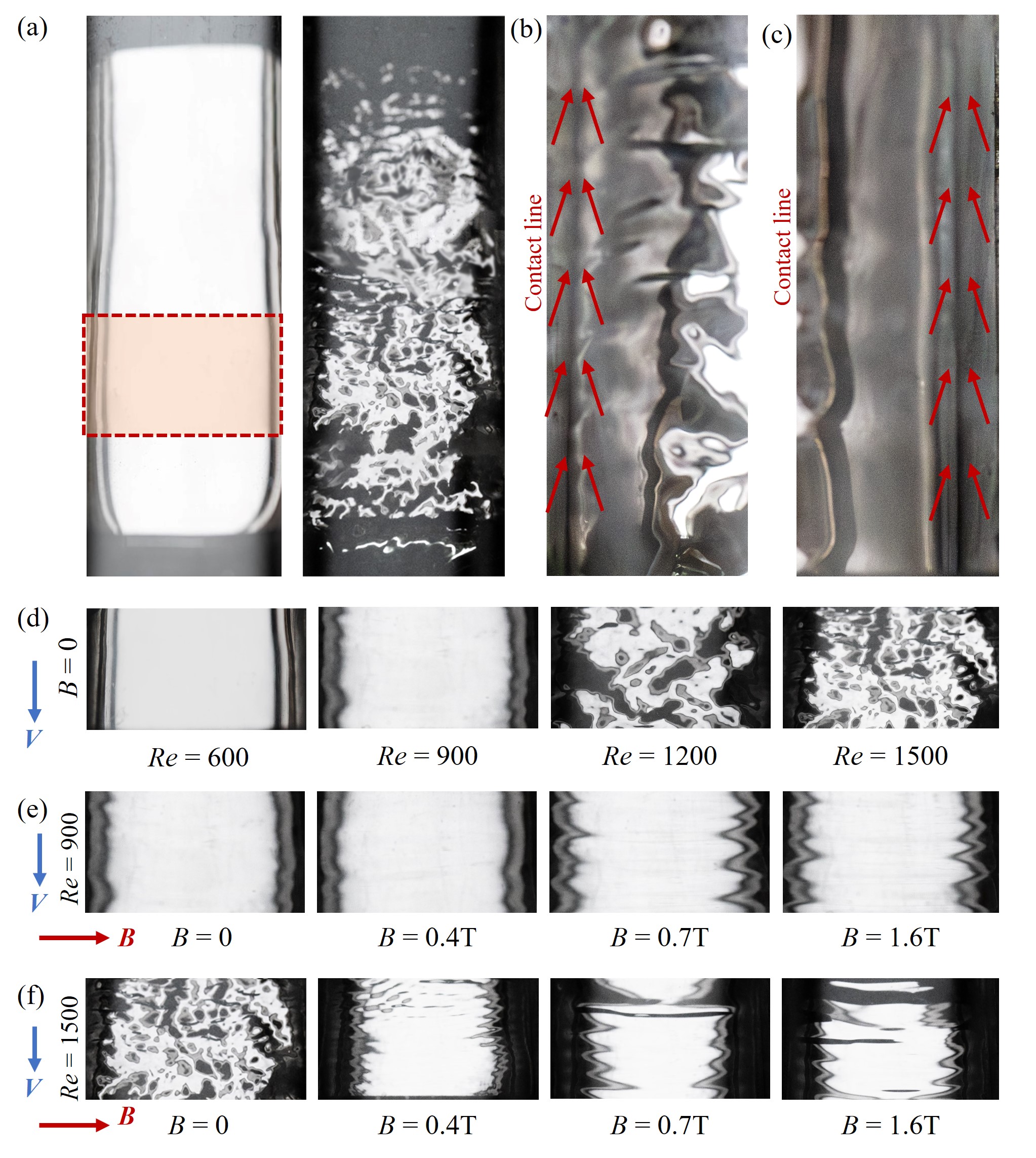}
\caption{Flow surface morphologies of film over MIFILM substrate, (a) both laminar and turbulent flows can cover the entire substrate. (b) is a close-up of the contact line of the liquid and the sidewall at \textit{Re} = 1500, \textit{B} = 0. (c) is a close-up of the contact line at \textit{Re} = 1500, \textit{B} = 1.6 T. Focusing on the area within the red box in (a), (d) shows different Re conditions with no magnetic field, (e) presents varying \textit{B} at \textit{Re} = 900, and (f) illustrates different \textit{B} at \textit{Re} = 1500.}\label{f9}
\end{figure}

Situations are different on the MIFILM substrate's solid-liquid composite surface. Due to the enhanced wettability, as the flow rate increases from the minimum stable flow rate of the pump \textit{Q} = 45 ml/min (\textit{Re} = 33.75) to \textit{Q} = 2000 ml/min (\textit{Re} = 1500), the liquid metal can spread over the entire substrate. No dewetting case is observed during the experiments.

\Fref{f9}(a) shows that the flow maintains a full-span spreading whether the flow is laminar or turbulent. \Fref{f9}(b) and (c) show that the liquid-solid contact lines are fixed on the sidewalls in turbulence and strong magnetic fields. \Fref{f9}(d), (e), and (f) provide more detailed flow characteristics taken from the mid-lower stream positions (\textit{x} = 120 - 170 mm), with parameter ranges of \textit{Re} = 600 - 1500, \textit{Ha} = 0 - 130.

\begin{figure}
\centering
\includegraphics[width=0.7\linewidth]{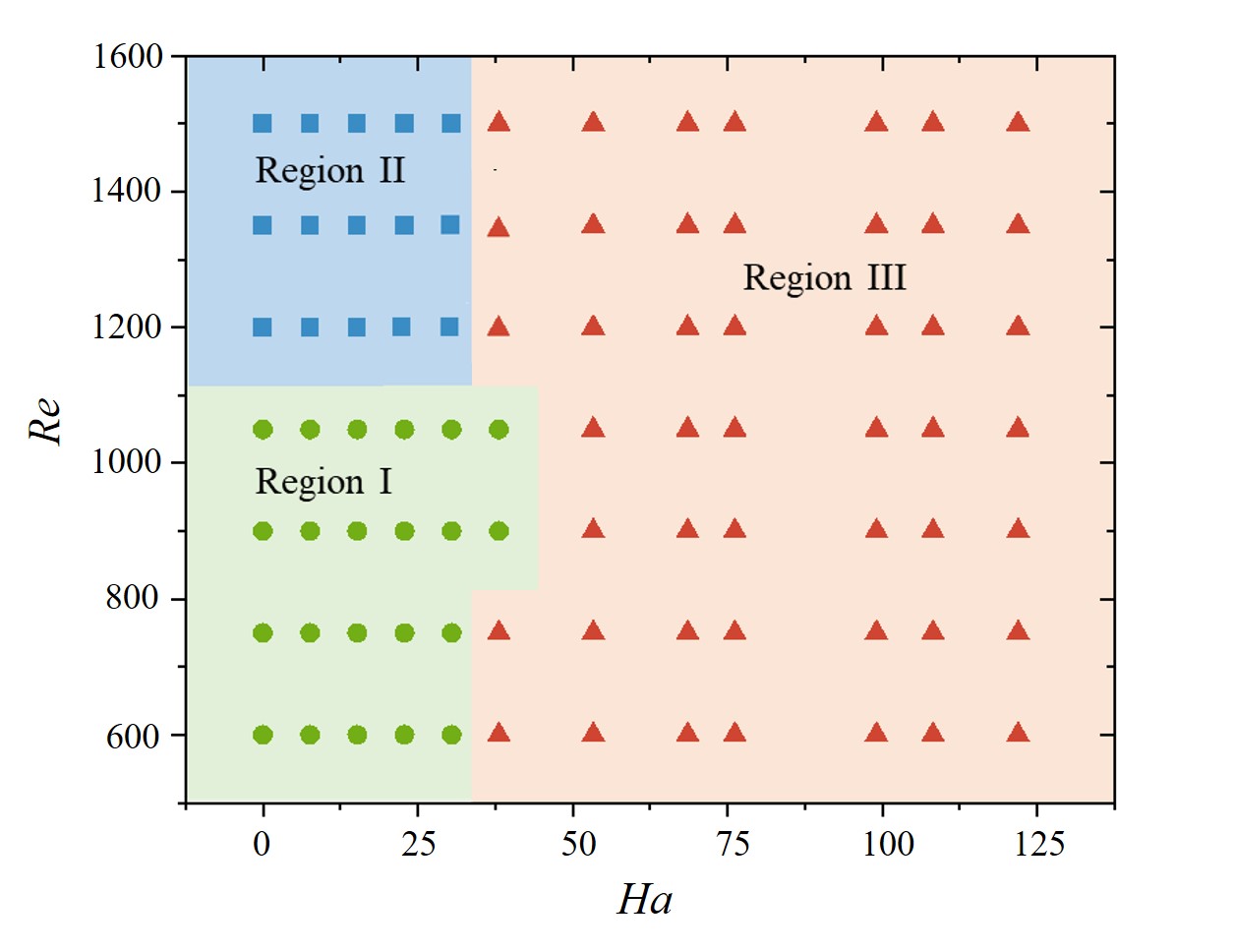}
\caption{Diagram of different flow regions, with Region I representing the quasi-laminar state, Region II indicating the three-dimensional fluctuation state, and Region III denoting the quasi-two-dimensional fluctuation state. }\label{f10}
\end{figure}

At \textit{Re} = 1500, the flow surface exhibits chaotic three-dimensional fluctuations in all directions. After the magnetic field is applied, these fluctuations become more orderly, waves parallel to the magnetic field are suppressed, the waves are mainly along the flow direction, and the flow tends to be quasi-two-dimensional. This observation fits classical theories that reported the influence of magnetic fields on turbulence, specifically that vortices perpendicular to the magnetic field direction are suppressed, while those parallel to the magnetic field are retained \cite{sommeria1982and,davidson1995magnetic}. 

At low Reynolds numbers, no significant waves are observed in any direction. However, significant streamwise waves are observed after applying the magnetic field, demonstrating a wave-generating effect of the magnetic field. Due to the conversion of kinetic energy, the vortices parallel to the magnetic field (shown as streamwise waves) are enhanced \cite{fauve1984chaotic}. As the magnetic field strength increases, these streamwise waves are amplified and detectable.

\Fref{f10} classifies the flow region diagrams on a \textit{Ha}-\textit{Re} map according to the surface waving patterns. At low Reynolds numbers and low Hartmann numbers (Region I), surface waves are nearly undetectable. In high-Reynolds-number and low-Hartmann-number regimes (Region II), turbulent flow generates complex three-dimensional wave patterns in all directions. In cases of high Hartmann numbers (Region III), the strong rearrangement effect of the magnetic field causes two-dimensional waves aligned with the flow direction.

\begin{figure*}
\includegraphics[width=1\linewidth]{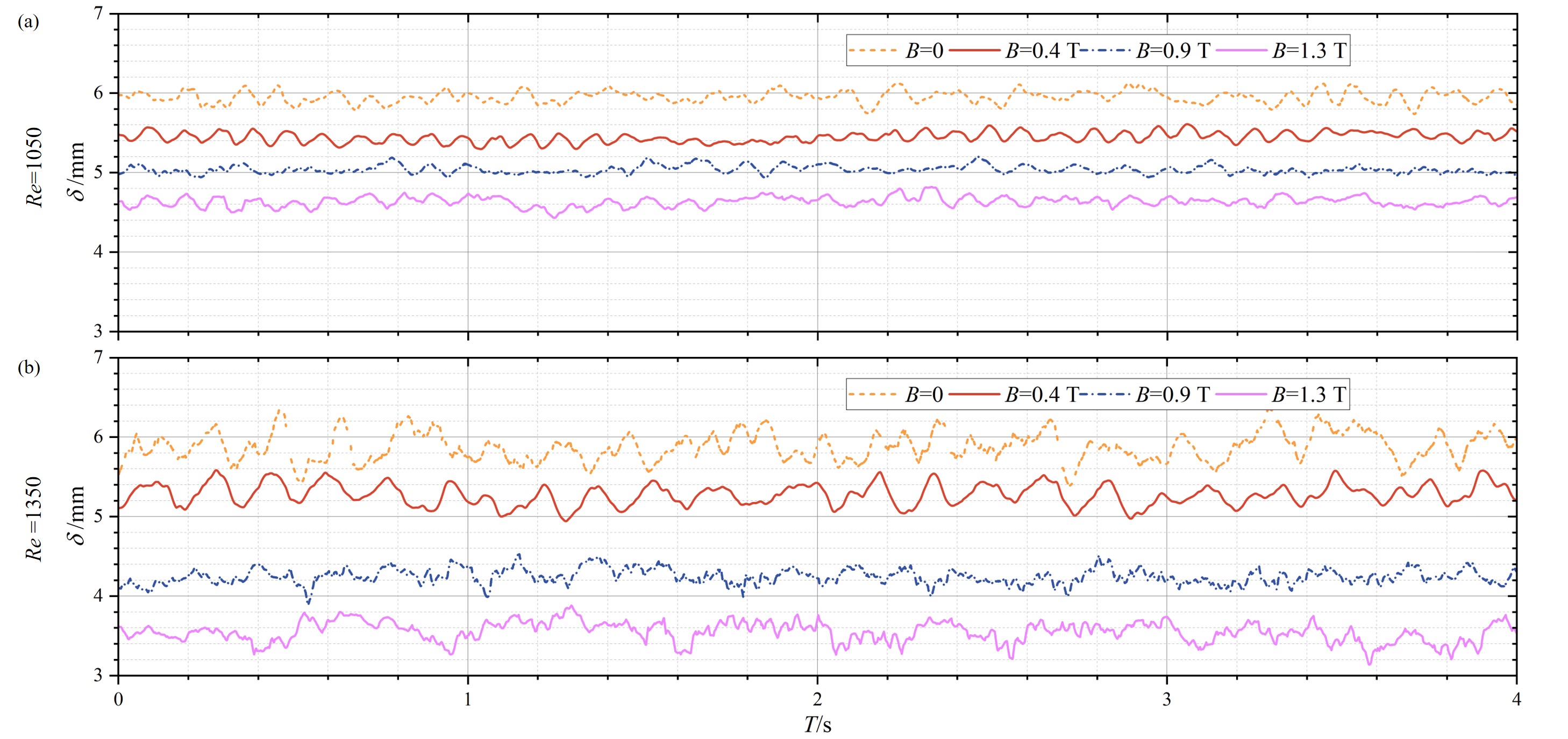}
\caption{The height variations and fluctuation images of the liquid at the downstream measurement point (\textit{x} = 160 mm) of different magnetic field strengths at (a) \textit{Re} = 1050 and (b) \textit{Re} = 1350. }\label{f11}
\end{figure*}

Additionally, using the CCS allows for more precise measurements of the liquid surface height changes. \Fref{f11} illustrates the height fluctuations over time at a fixed point downstream (\textit{x} = 160 mm). It can be observed that low flow rate results in a smoother surface with reduced fluctuations. With the increase of magnetic field intensity, the amplitude of the fluctuations decreases first and then increases. The aforementioned surface photos taken by the camera show that the direction of enhanced waves is mainly streamwise. Together, they indicate that the 2D effect of the magnetic field not only suppresses transverse waves but also enhances streamwise waves.

The experiment demonstrates the spreading behavior of the insulating MIFILM substrate in a transverse magnetic field, which reduces the minimum flow rate required for liquid spreading compared to flat substrates. The transition to quasi-2D flow is observed at high flow rates, similar to that on classical flat substrates. Moreover, at low flow rates, the wave generation effect is observed. 

\subsection{Effects of transverse magnetic fields }

\begin{figure}
\centering
\includegraphics[width=0.6\linewidth]{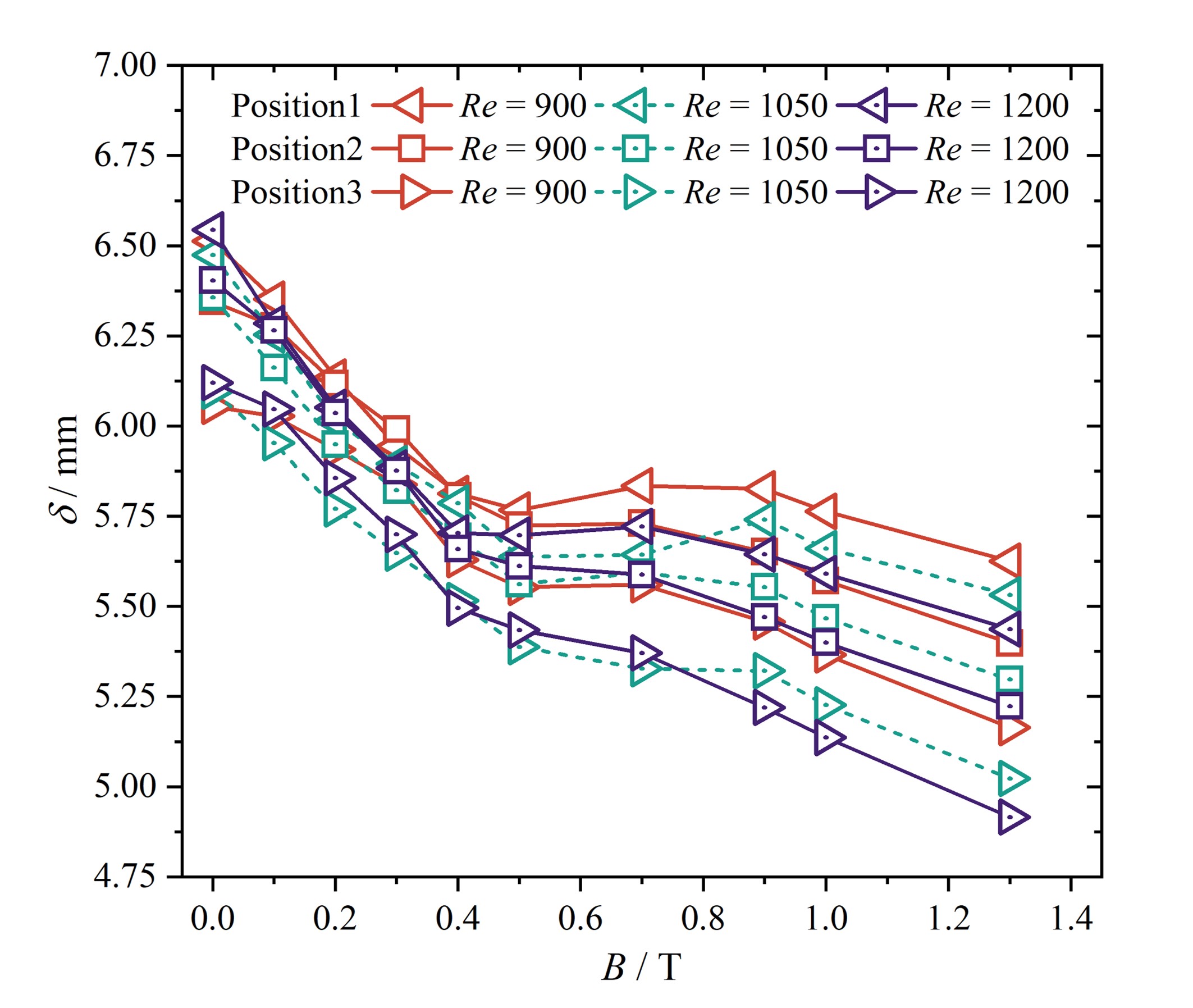}
\caption{Variation of film thickness at different Reynolds numbers and magnetic field strengths at three positions (Position 1: \textit{x} = 40 mm, Position 2: \textit{x} = 100 mm, Position 3: \textit{x} = 160 mm).}\label{f12}
\end{figure}

\Fref{f12} shows the variation of the average film thickness at three positions concerning magnetic field strength under different \textit{Re}. In the absence of a magnetic field, the film thickness gradually decreases from upstream to downstream, while the effect of the magnetic field further amplifies the film thickness difference between the upstream and downstream regions. This is similar to the film thickness distribution obtained from experimental measurements and simulations in \cite{sun2023magnetohydrodynamics}.

In the absence of a magnetic field, a higher Reynolds number generally corresponds to a thicker film thickness. When a magnetic field is applied, the film thickness does not increase but decreases. This reduction in film thickness is hypothesized to arise from the aforementioned 2D effects of the magnet field, as the main flow direction gains more kinetic energy from other directions. When the magnetic field strength reaches approximately 0.5 T, the trend of decreasing film thickness slows down, corresponding to a critical Hartmann number \textit{Ha\textsubscript{cr}} = 40. The overall variation trend is similar across the entire field, possibly limited by the flow length (200 mm in the flow direction).

\begin{figure}
\centering
\includegraphics[width=0.6\linewidth]{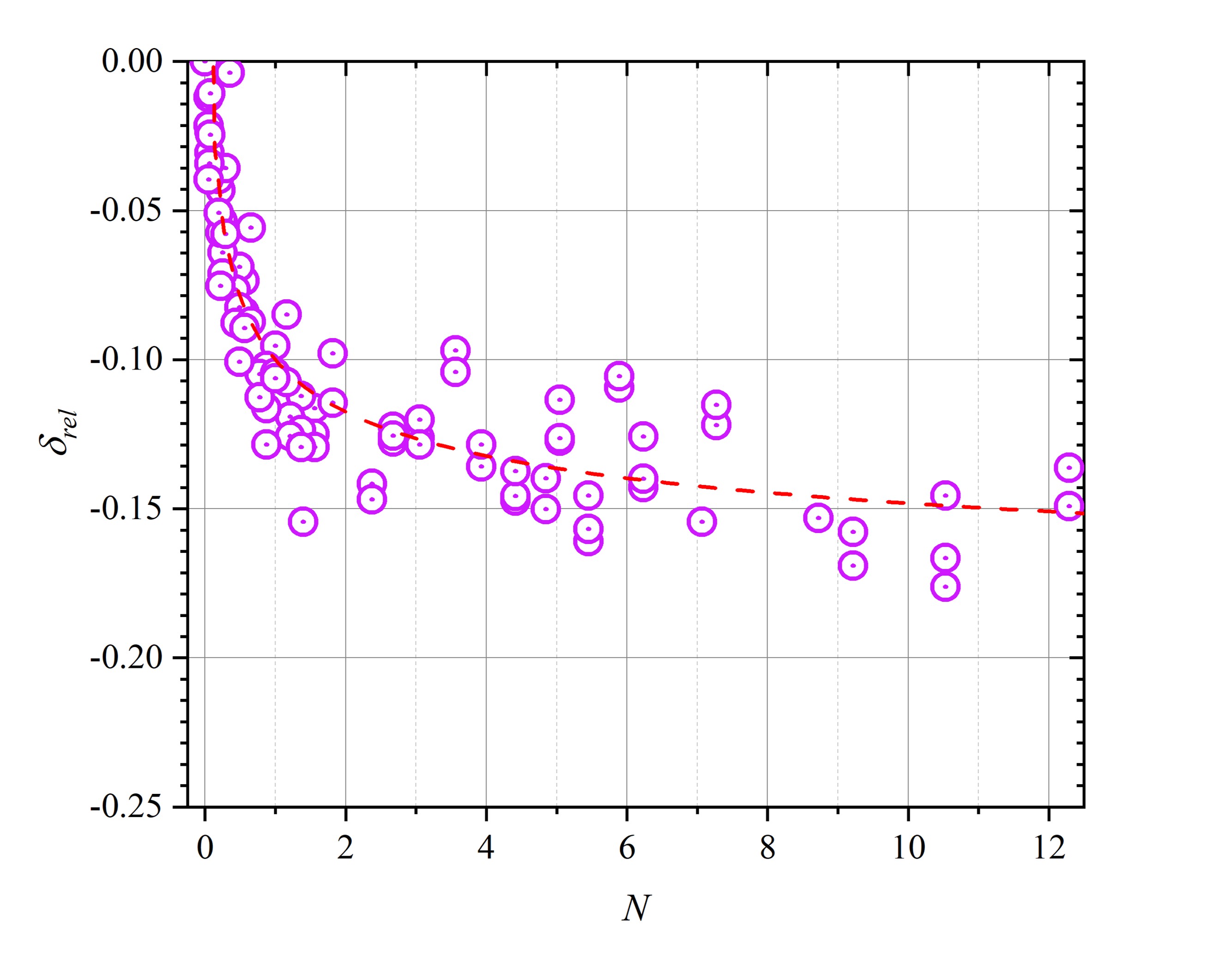}
\caption{The variation in ${\delta _{rel}}$ as a function of \textit{N} during the flow of pure liquid metal on the MIFILM substrate. }\label{f13}
\end{figure}

\Fref{f13} categorizes the change in film thickness concerning the Stuart number. The dimensionless relative change in film thickness is defined as:
\begin{eqnarray}
    {\delta }_{rel}={(\delta -{{\delta }_{0}})}/{{\delta }_{0}} ,
\end{eqnarray}
where ${\delta }$ is the average film thickness and ${\delta _{0}}$ represents the thickness with no magnetic field at the same flow rate as a reference for comparison. 

On the insulating MIFILM substrate, the dimensionless film thickness variation generally decreases as the Stuart number increases, and the rate of decline in film thickness gradually slows down. The fitted curve (shown as the dashed line in \fref{f13}) is:
\begin{eqnarray}
    \delta _{rel}=0.101{{N}^{-0.25}}-0.203 .
\end{eqnarray}

In the form of a scaling law, it can be expressed as ${{\delta }_{rel}}\propto {{N}^{-0.25}}$. Alternatively, using the critical Stuart number $\textit{N}_{cr}$ = 1 as a boundary, when \textit{N} $<$ 1, the film thickness decrease is rapid. When \textit{N} $>$ 1, the decrease noticeably slows down and stabilizes, consistent with computer simulations reported in \cite{pan2025magnetohydrodynamic}.

\begin{figure}
\centering \
  \includegraphics[width=0.6\linewidth]{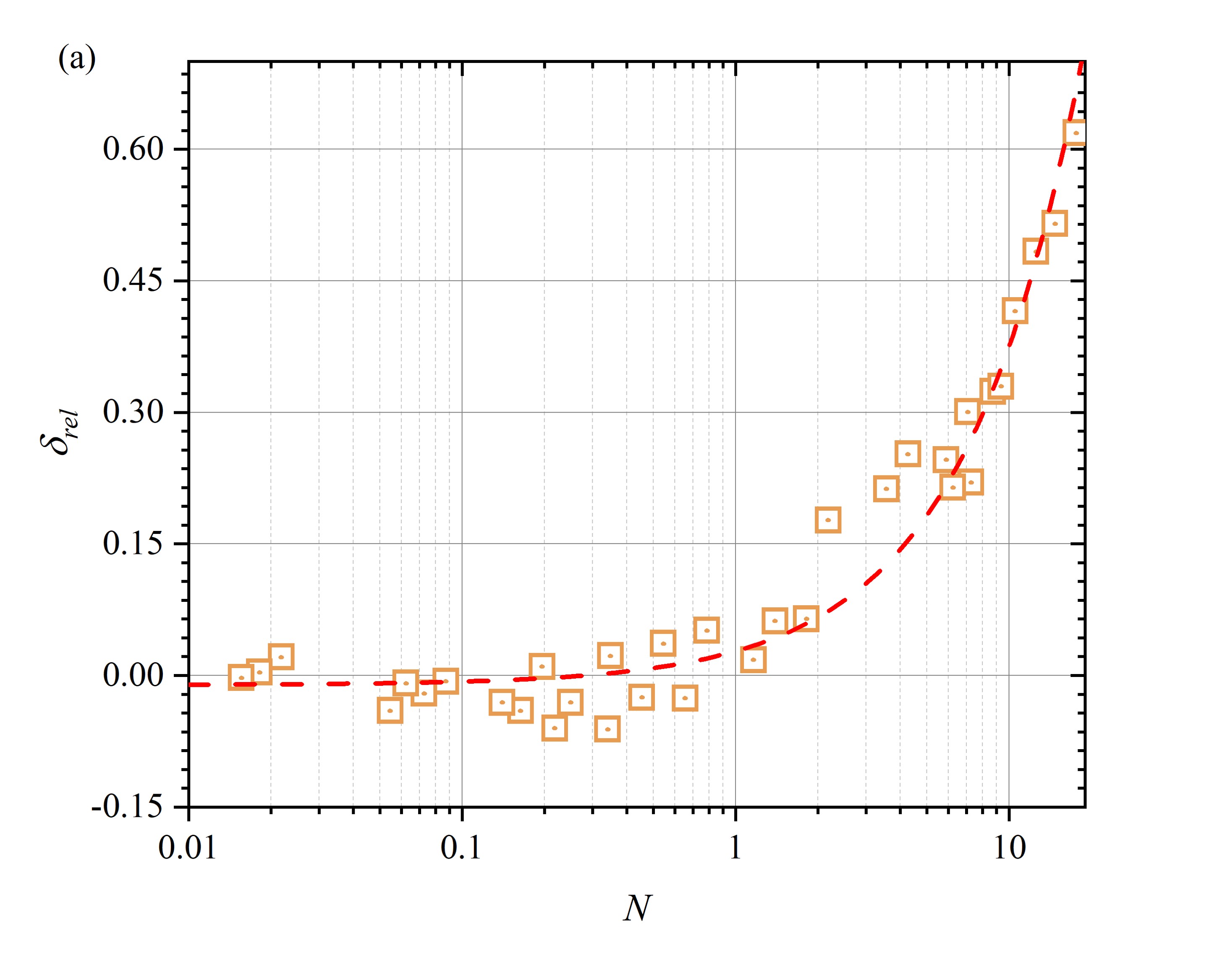}
 \includegraphics[width=0.6\linewidth]{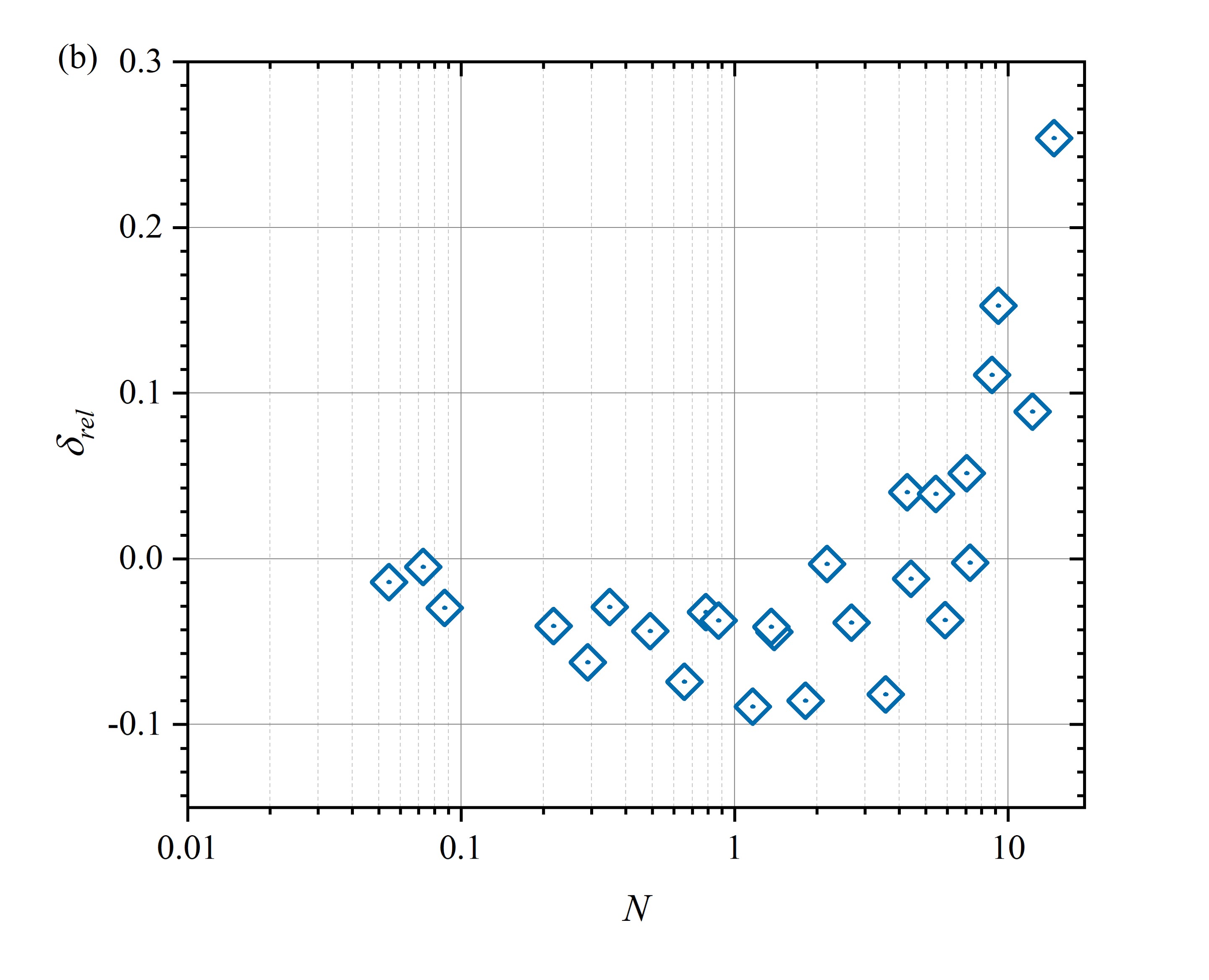}
\caption{The variation in ${\delta _{rel}}$ as a function of \textit{N} for the flow film of oxidized liquid metal on flat substrates with (a) conductive wall and (b) insulating wall.}\label{f14}
\end{figure}

In the experiments with conductive wall conditions at high flow rates described in reference \cite{yang2020magnetohydrodynamic}, it was observed that the film thickness showed no significant change until a certain critical Hartmann number \textit{Ha\textsubscript{cr}} was reached. Then the film thickness increased rapidly. Due to the notable differences in film thickness variation, we conducted a comparative investigation on flat substrates. The liquid metal flow can spread through oxide skin adhesion under slight oxidation conditions. The operational method is to open two valves of the liquid metal loop. One ventilates the ambient environment, and the other introduces argon gas to prevent oxide blockage formation. Two cases are tested: an insulating wall (sidewalls and substrate made entirely of resin) condition and a conductive wall (made entirely of stainless steel) condition. The results are compared with those of the current insulating MIFILM substrate.

On the conductive flat substrate, the ${\delta _{rel}}$ generally shows an increasing trend with the increase in \textit{N}, and the increase rate of ${\delta }_{rel}$ becomes progressively faster, which can be expressed as the scaling law ${{\delta }_{rel}}\propto N$ \cite{yang2020magnetohydrodynamic}. Our results are also consistent with this relation. The fitted curve (shown as the dashed line in \fref{f14}(a)) is:
\begin{eqnarray}
    {{\delta }_{rel}}=0.0386N-0.0109 .
\end{eqnarray}

Furthermore, the critical Stuart number \textit{N\textsubscript{c}}\textit{\textsubscript{r}} = 1 is the dividing line, when \textit{N} $<$ \textit{N\textsubscript{c}}\textit{\textsubscript{r}} it can be considered that the relative film thickness does not change with \textit{N}, while \textit{N} $>$ \textit{N\textsubscript{c}}\textit{\textsubscript{r}} the relative film thickness increases with the increase of \textit{N}, as shown in \fref{f14}(a).

For the experiments of the insulating flat substrate, the relative film thickness decreases first and then increases with the increase of \textit{N}, and the critical Stuart number \textit{N\textsubscript{c}}\textit{\textsubscript{r}} = 1 is taken as the dividing line, shown in \fref{f14}(b). This means that the evolution of the relative film thickness on the insulating flat substrate is between the insulating MIFILM and the conductive flat substrate.

In the parameter range of the low flow rate experiments presented in this study, the influence of the magnetic field on the film flow can be considered as a combination of the two-dimensional effect and the MHD damping effect. And there is a competitive mechanism between the two effects. For all three experimental conditions, below the critical Stuart number, the flow gains more kinetic energy in the flow direction due to the two-dimensional effect, where the Lorentz force’s magnetic damping effect is relatively weak. However, when the magnetic field exceeds the critical Stuart number, the influence of MHD resistance becomes more significant, while the kinetic energy conversion approaches completion, causing its effect to decrease.

What’s more, the conductive walls allow current to flow along the wall. Due to the presence of this enhanced current, the flow of liquid metal is subject to a more significant drag, thus the MHD damping effect dominates. In contrast, the insulating wall cannot conduct electricity, so the current is concentrated in the flow area inside the liquid. Since no additional current is generated at the wall, the Lorentz force caused by the magnetic field is relatively weaker. In this case, the two-dimensional effect of velocity is more obvious than that of magnetic damping. For the oxidized flow, we speculate that the solid oxide layer adheres to the side wall, resulting in a certain degree of conductivity with its flow characteristics between conductive and insulating.
\begin{figure}
\centering
\includegraphics[width=0.6\linewidth]{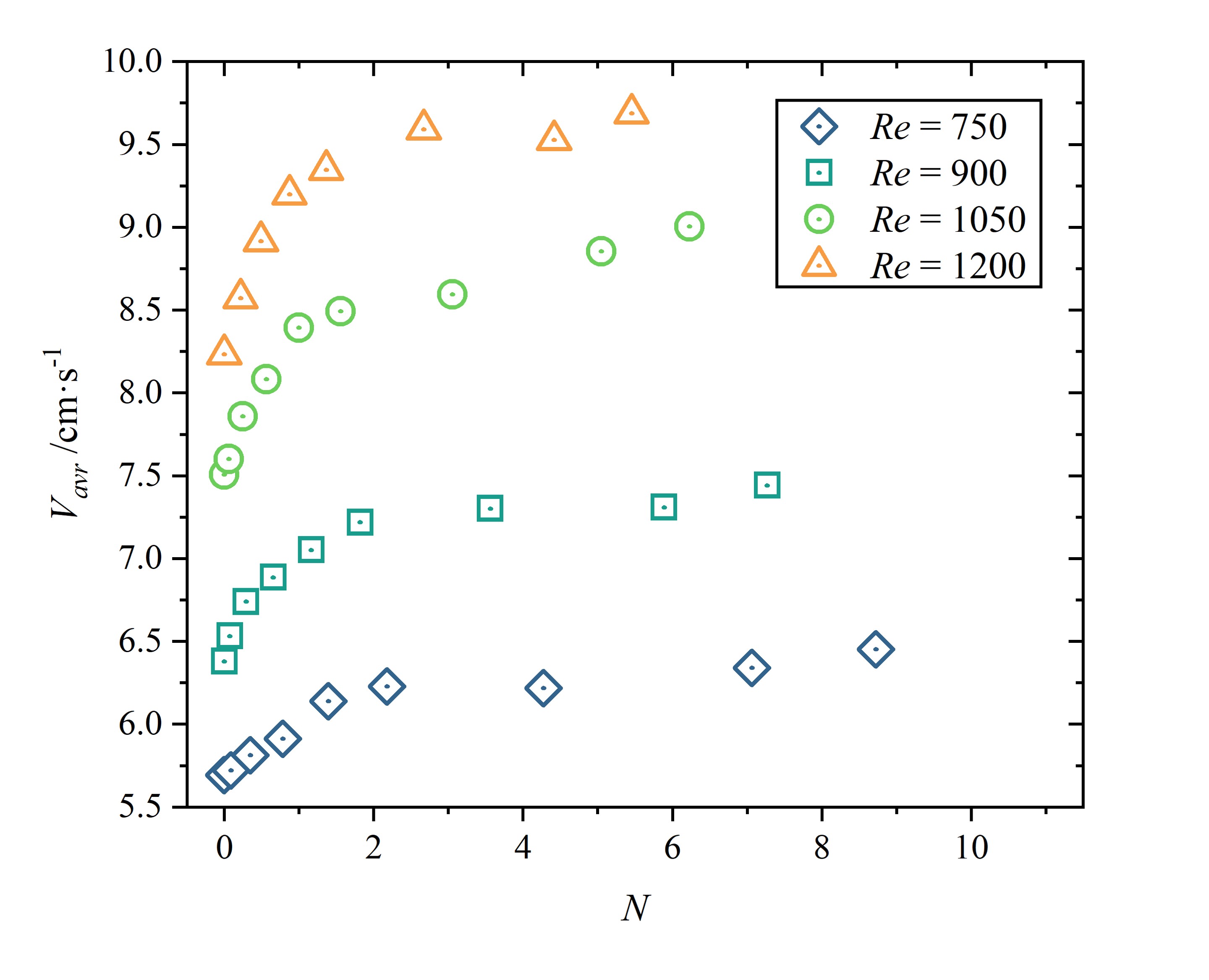}
\caption{Average velocity calculated at the downstream measurement point (\textit{x} = 160 mm) based on flow rate and film thickness. }\label{f15}
\end{figure}

This explains three different behaviors observed in the experiments: with the increase of \textit{N}, the relative film thickness of pure GaInSn on the insulating MIFILM substrate decreases first and then becomes stable, while the film thickness of oxidized GaInSn on the conductive flat substrate stabilizes first and then increases. On the partially conductive flat substrate, the film thickness of oxidized GaInSn decreased first and then increased. 

When the flow rate is constant, if the flow width is unchanged, the film thickness is inversely proportional to the flow velocity. Using the measured average film thickness, the averaged flow velocity on the MIFILM substrate is calculated: ${{V}_{avr}}={Q}/{w\delta }\;$. Downstream velocities at the measurement station (\textit{x} = 160 mm) are shown in \fref{f15}. The flow speed increases with magnetic field strength, and shows a gradually reduced growth rate. MIFILM substrate exhibits better flow spreading performance than flat substrates in strong transverse magnetic fields, requiring a driving speed on the order of only centimeters per second. In addition to spreading advantages, the magnetic resistance of the flow is also lower on MIFILM substrate. 

\section{Conclusion}\label{s4}
To address the spreading issues of liquid metals, this study proposes a novel substrate design strategy that utilizes pre-filled micropillar arrays to achieve uniform film flow of liquid metal on non-wetting surfaces without chemical modifications. 

By constructing a triangular array of thumbtack-shaped reentry micropillars on the non-wetting surface, and using a vacuum pre-filling method to press liquid metal into the inter-pillar gaps, MIFILM substrate with a solid-liquid composite surface is created. MIFILM reduces the contact angle of the liquid metal from 140° to approximately 20°, and transitions the substrate from hydrophobic to hydrophilic. By comparing the liquid metal spreading characteristics on the flat and MIFILM substrate, the results showed that the metal flow on MIFILM substrate exhibits stable and complete spreading under all experimental conditions, including at low or high flow rates, and under strong magnetic fields. 

The flow patterns under transverse magnetic fields were investigated. For laminar flows, magnetic fields induce streamwise waves. For turbulent flows, magnetic fields suppress transverse waves and enhance streamwise waves. Experimental results also indicated that the liquid film on insulating MIFILM substrates thins with increasing magnetic field before reaching a stable thickness at higher fields because of a combined influence of the magnetic field's MHD drag and 2D effect. 

Compared to conventional flat substrates, MIFILM substrates can enable spreading at substantially lower flow rates while maintaining smooth flow characteristics and minimizing MHD damping effects. This combination of stable, well-ordered flow patterns and minimal flow resistance makes MIFILM a particularly suitable substrate for flowing liquid metal PFCs.

On this basis, the advancement of metal 3D-printing technology offers the potential to fabricate the proposed microstructured substrates with steel, molybdenum, or tungsten. These metal substrates can be filled with liquid lithium to create full-metal MIFILM PFCs for functional performance testing in plasma devices. Essential tests evaluating thermal protection, plasma interactions, and evaporation are required for future applications in Tokamak devices.

\ack{}
This research is supported by the National Key R\&D Program of China (No. 2022YFE03130000) and the National Natural Science Foundation of China (No. 12372266).

\section*{References}

\bibliographystyle{unsrt}
\bibliography{main}

\end{document}